%% file: main.tex
\documentclass[letterpaper,twocolumn,10pt]{article}
\usepackage{usenix2025_SOUPS}
\usepackage{tikz}
\usepackage{amsmath}
\usepackage{graphicx}
\usepackage{xcolor}
\usepackage{soul}
\usepackage{enumitem}
\usepackage{booktabs}
\usepackage{comment}
\usepackage{tcolorbox}
\usepackage{xurl}
\usepackage{caption}

\newcommand{\num}{135 }

\begin{document}

\date{}

\title{\Large \bf 
Measuring NIST Authentication Standards Compliance\\by Higher Education Institutions}

\def\plainauthor{Noah Apthorpe, Boen Beavers, Yan Shvartzshnaider, Brett Frischmann}

\author{
{\rm Noah Apthorpe}\\
Colgate University
\and
{\rm Boen Beavers}\\
Colgate University
\and
{\rm Yan Shvartzshnaider}\\
York University\\
\and
{\rm Brett Frischmann}\\
Villanova University
} 

\maketitle
\thecopyright

\begin{abstract}
\input{sections/abstract}
\end{abstract}

\input{sections/introduction}
\input{sections/background}
\input{sections/related}
\input{sections/methods}
\input{sections/results}
\input{sections/discussion}
\input{sections/conclusion}

\section*{Acknowledgments}
We thank Julia Tuck for her contributions. We also thank
Max Alphonso,
Pierce Anthony,
Aly Bannister,
William Beckhorn,
Ryan Brackett,
Matthew Burack,
Onil Carrion,
Jackson Carter,
Sophia Cecchin,
Danny Chu,
Brian Douglas,
Lily Ellis,
Kate Gallagher,
Julia Goosay,
Jasper Gough,
Alex Greene,
Carly Grizzaffi,
Nikki Izversky,
Nephileen Kattel,
Natalie McCall,
Paige Mizutani,
Aidan Murnane,
Alex Tauber,
Gray Theodore, 
and
Blake Wang.
This research was
supported by the Colgate University Faculty Research Council. Yan Shvartzshnaider is supported in part by funding from the Social Sciences and Humanities Research Council of Canada.

\bibliographystyle{plain}
\bibliography{main}

\end{document}

%% file: sections/abstract.tex
Technical standards are a longstanding method of communicating best practice recommendations based on expert consensus. Cybersecurity standards are particularly important for informing policies that protect critical systems and sensitive data. Measuring standards compliance is therefore essential to identify vulnerabilities arising from outdated policies and to determine whether expert advice has effectively diffused to practitioners.
In this paper, we examine the authentication policies of a diverse set of \num colleges and universities in the United States and Canada to determine compliance with four standards from NIST Special Publication 800-63 \textit{Digital Identity Guidelines}. 
We find widespread, but not universal, deployment of multi-factor authentication across institutions. We also find prevalent outdated use of password expiration, password composition rules, and knowledge-based authentication. 
These results support further investment and research into incentive structures for standards compliance and the diffusion of expert guidance to practitioners.

%% file: sections/introduction.tex
\section{Introduction}
\label{sec:introduction}
The 2004 National Institute of Standards and Technology (NIST) Special Publication (SP) 800-63 \textit{Electronic Authentication Guideline}~\cite{800-63-v2004}  advised users to secure their accounts with complex passwords composed of random characters, capital letters, and numbers, and to change their passwords regularly~\cite{mcmillan2017comptia}. Federal agencies, corporations, and universities largely followed this advice by enforcing password complexity rules and regular password expiration. However, the expert consensus on authentication has changed dramatically in the intervening years as the field has broadened beyond computational cybersecurity to include human factors, usability, and other knowledge sources \cite{frischmann2023common}.
Recognizing flaws and limitations of the 2004 publication, NIST published major updates in 2011 \cite{800-63-1}, 2013 \cite{800-63-2}, and 2017 \cite{800-63-3, 800-63A, 800-63B, 800-63C}. Some of these updates reflected advances in authentication technologies; for example, multi-factor authentication is now strongly recommended. Others represented 180-degree changes from previous standards based on evolving expert knowledge; for example, forced password composition rules, regular password expiration, and knowledge-based authentication are now strongly discouraged~\cite{800-63B}.

For several reasons, these specific updates to NIST's authentication standards,
codified in the 2017 NIST SP 800-63 \textit{Digital Identity Guidelines}~\cite{800-63-3, 800-63A, 800-63B, 800-63C}, create an ideal case study for examining how well expert cybersecurity knowledge is reaching practitioners responsible for system implementation, maintenance, and governance.
First, the stark differences between the current and previous standards make it easy to detect stale institutional policies.
Second, the current standards have been in place since 2017, meaning that all institutions have had ample time to update their policies if they are going to do so.
Finally, the practices recommended by the current standards are widely endorsed by academic research (Section~\ref{sec:related}) and by other industry and government bodies \cite{FIDO2020NISTComments, Kantara2018TrustMark, 800-63-FAQ, cse2018itsp}. This means that noncompliance with these standards is more likely indicative of a general disconnect between expert recommendations and institutional policies rather than an inapplicability of the standards to a particular institution.

In this paper, we 
examine the authentication policies of a diverse set of \num colleges and universities in the United States and Canada, including public, private, R1, liberal arts, historically Black, and regional institutions. We review these institutions’ online documentation and check for compliance with the four standards from NIST SP 800-63 mentioned above regarding (1) multi-factor authentication, (2) password expiration, (3) password composition rules, and (4) knowledge-based authentication. 
We observe widespread, but not universal, deployment of multi-factor authentication across the institutions we examine. 
We also observe that password expiration, specific password composition rules, and knowledge-based authentication are still common contrary to NIST's guidelines. 

Prior work by Lee et al.~\cite{lee2022password} and Hall et al.~\cite{hall2023empirical} has found low compliance with updated NIST authentication standards across the tech industry; however, ours is the first study to date that has examined compliance by higher education institutions specifically. 
Focused attention on this sector is relevant, because higher education institutions have been increasingly targeted by cyberattacks attempting to compromise authentication credentials~\cite{fbi2022academiccredentials}.
The widespread deployment of multi-factor authentication we observe is highly encouraging 
given that multi-factor authentication substantially increases the difficulty of credential-based attacks. Nevertheless, the outdated password and knowledge-based authentication policies we observe 
suggest a need for
further examination of incentive structures and more research into knowledge diffusion barriers that result in noncompliant and insecure policies.

The rest of this paper is structured as follows. Section~\ref{sec:background} details the specific standards from NIST SP 800-63 that we focus on in this paper, including NIST's rationale for updating them from pre-2017 versions. 
Section~\ref{sec:related} reviews related academic research supporting the standards updates, measuring compliance with  NIST SP 800-63 in other contexts, and about authentication practices in higher education generally. 
Section~\ref{sec:methods} describes our data collection and analysis methods, including research ethics and limitations. 
Section~\ref{sec:results} presents our results, including rates of compliance and noncompliance across institutional categories.  
Section~\ref{sec:discussion} discusses the implications of these results, emphasizing their significance for the usable security community. 
Section~\ref{sec:conclusion} concludes.

%% file: sections/background.tex
\section{Background and Research Questions}
\label{sec:background}

NIST SP 800-63 \textit{Digital Identity Guidelines} was published in 2017 as a four-volume document suite (SP 800-63-3 \cite{800-63-3}, SP 800-63A \cite{800-63A}, SP 800-63B \cite{800-63B}, and SP 800-63C \cite{800-63C}).  
This was an update to SP 800-63-2 (2013) \cite{800-63-2} and as part of a series of authentication standards documents including SP 800-63-1 (2011) \cite{800-63-1} and SP 800-63 (2004) \cite{800-63-v2004}.  The 2017 version was developed in collaboration with the community, during which it received over 1,400 comments in a draft period from 2016–2017.  
NIST SP 800-63 provides ``an overview of identity frameworks; using authenticators, credentials, and assertions in a digital system; and a risk-based process to select assurance levels’’ \cite{800-63-3}. The digital authentication standards presented in NIST SP 800-63 are extensive and cover practices across the technical stack and of differing relevance to different types of organizations. 

We selected four standards from NIST SP 800-63 
to investigate for this study. We chose these specific standards because they form an ideal case study of the diffusion of expert cybersecurity knowledge to practitioners for reasons presented in Section~\ref{sec:introduction} and summarized here:
First, clear differences between the current and previous versions of these standards make it easy for us to detect outdated institutional policies.
Second, these standards have been in place since 2017, providing institutions ample time to update their policies.
Third, these standards are widely supported by academic research and other government bodies, meaning that noncompliance is more likely due to a general disconnect between expert recommendations and institutional policies than due to conflicting expert advice.

The following subsections describe each of the standards we investigate in detail.

\subsection{Multi-factor Authentication}

Multi-factor authentication (MFA) is well understood to be the best practice for digital authentication. Requiring two or more different forms of authentication, especially the combination of a memorized secret (e.g., a password) and a possession-based authenticator (i.e., ``something you have''), makes account compromise significantly more challenging for would-be adversaries. 

Pre-2017 versions of NIST SP 800-63 used a "levels of assurance" (LOA) model to determine what authentication procedures should be followed. The required level depended on the amount of risk posed by an authentication error. Level 1 (no risk) did not require or recommend MFA. Level 2 (moderate risk) recommended, but did not require, MFA. Only Levels 3 and 4 (high and very high risk) required MFA.

The 2017 version of NIST SP 800-63 changed to a more nuanced model with separate levels for identity assurance (IAL), authenticator assurance (AAL), and federation assurance (FAL). AAL 1 (low risk) recommends, but does not require, MFA. AAL 2 (moderate risk) requires MFA:
\begin{tcolorbox}[colframe=black,colback=white]
 {\small 
Authentication SHALL occur by the use of either a multi-factor authenticator or a combination of two single-factor authenticators. \cite{800-63B}
}
\end{tcolorbox}
\noindent AAL 3 (high risk) requires MFA with a hardware-based authenticator. The inclusion of a MFA requirement for situations where an authentication error poses ``moderate risk'' is a significant expansion of of the MFA requirement from the pre-2017 versions. Since authentication errors at higher education institutions pose at least ``moderate risk,'' we investigate whether or not the institutions examined in this paper require MFA. 

\subsection{Password Expiration}

All pre-2017 versions of NIST SP 800-63 recommended regular password expiration or cycling as a best practice. This received a 180-degree update in the 2017 version, as stated in
NIST SP 800-63B Section 5.1.1.2:

\begin{tcolorbox}[colframe=black,colback=white]
 {\small 
Verifiers SHOULD NOT require memorized secrets to be changed arbitrarily (e.g., periodically). However, verifiers SHALL force a change if there is evidence of compromise of the authenticator. \cite{800-63B}
}
\end{tcolorbox}

The previous wisdom held that the likelihood of password compromise increases over time and that regular password expiration is needed to reduce this risk.  
However, the NIST SP 800-63 FAQ provides the rationale for the new standard:

\begin{tcolorbox}[colframe=black,colback=white]
 {\small 
Users tend to choose weaker memorized secrets when they know that they will have to change them in the near future. When those changes do occur, they often select a secret that is similar to their old memorized secret by applying a set of common transformations such as increasing a number in the password. This practice provides a false sense of security if any of the previous secrets has been compromised since attackers can apply these same common transformations.  \cite{800-63-FAQ}
}
\end{tcolorbox}

We therefore investigate whether higher education institutions require or recommend regular password expiration or cycling for their affiliates.

\subsection{Password Composition Rules}
All pre-2017 versions of NIST SP 800-63 recommended password composition rules during account creation. The 2017 version of the standard makes a 180-degree change to this recommendation, as stated in
NIST SP 800-63B Section 5.1.1.2: 

\begin{tcolorbox}[colframe=black,colback=white]
 {\small 
Verifiers SHOULD NOT impose other composition rules (e.g., requiring mixtures of different character types or prohibiting consecutively repeated characters) for memorized secrets. \cite{800-63B}
}
\end{tcolorbox}

It is approaching common knowledge that all passwords “should” contain at least one symbol, number, and (often) mix of capital and lowercase letters. Most are familiar with password creation interfaces that enforce specific composition rules. 
However, these interfaces are all out of compliance with the current NIST standard. 

NIST SP 800-63B Appendix A.3 provides the rationale for this change:

\begin{tcolorbox}[colframe=black,colback=white]
 {\small 
Research has shown, however, that users respond in very predictable ways to the requirements imposed by composition rules [Policies]. For example, a user that might have chosen “password” as their password would be relatively likely to choose “Password1” if required to include an uppercase letter and a number, or “Password1!” if a symbol is also required. 
\cite{800-63B}
}
\end{tcolorbox}

We therefore investigate whether higher education institutions require or recommend specific composition rules as a part of their password creation processes. 

\subsection{Knowledge-based Authentication}
All pre-2017 versions of NIST SP 800-63 recommended knowledge-based authentication (KBA) as a component of a strong authentication posture. The 2017 version makes a 180-degree change to this recommendation, as stated in the NIST SP 800-63 FAQ:

\begin{tcolorbox}[colframe=black,colback=white]
 {\small 
Knowledge-based authentication (KBA), sometimes referred to as “security questions”, is no longer recognized as an acceptable authenticator by SP 800-63. This was formerly permitted and referred to as a “pre-registered knowledge token” in SP 800-63-2 and earlier editions. The ease with which an attacker can discover the answers to many KBA questions, and relatively small number of possible choices for many of them, cause KBA to have an unacceptably high risk of successful use by an attacker. \cite{800-63-FAQ}
}
\end{tcolorbox}

Despite this change, the use of KBA for digital authentication is still widespread. Most laypeople are familiar with the need to provide answers to questions such as “What was the model of your first car?” “What was your grandmother’s maiden name?” and  “What is your favorite winter sport?” (as well as a wide variety of others) during account creation and authentication. The use of such questions is no longer acceptable.
We therefore investigate whether higher education institutions require or recommend KBA during account creation, login, and/or recovery. 

\subsection{Canadian Standards}
The NIST standards we investigate have direct analogues in Canadian standards. For example, the \textit{Password Guidance} published by the Treasury Board of Canada Secretariat recommends that system owners  ``use two-factor authentication (2FA),'' ``eliminate password expiry,'' and ``disable or reduce complexity policies''~\cite{GCPasswordGuidance}.
The congruence of these standards supports our inclusion of 20 Canadian universities in this study.

%% file: sections/related.tex
\section{Related Work}
\label{sec:related}

The rapid rise in computational capabilities and credential leaks that allow attackers to crack a large number of passwords have forced researchers and practitioners to re-examine authentication policies, including multi-factor authentication, forced password expiration, password composition rules, and knowledge-based authentication. Scholars have also examined the actual behavior of individuals and companies regarding these and other authentication practices~\cite{da2022cyber, kumar2017no, bhagavatula2022adulthood, moh2024understanding, ray2021older, pearman2019people}; however, no prior study has compared authentication policies of higher education institutions to current industry standards as broadly as in this work.

\paragraph{Multi-factor Authentication.} 
Adoption of multi-factor authentication (MFA) has accelerated in recent years, with online services from all sectors, including higher education~\cite{weidman2017like} deploying optional or required MFA during login. In 2018, Colnago et al.~\cite{colnago2018s} studied user opinions about MFA at Carnegie Mellon University. Users found MFA ``annoying, but fairly easy to use, and believed it made their accounts more secure.'' In 2022, Arnold et al.~\cite{arnold2022emotional} similarly found that due to the ``time sensitive nature of many tasks that required MFA, university students are likely to experience strong negative emotions towards MFA that drastically lower their perceptions of its utility and usability,'' but that these emotions could be offset by an increased perception of security provided by MFA. These findings are corroborated by Dutson et al.~\cite{dutson2019don}, adding to a body of literature supporting widespread deployment of MFA at educational institutions.

\paragraph{Password Expiration.}
Several prior efforts provided early evidence about the relative ineffectiveness of password expiration policies.  In 2010, Zhang et al.~\cite{zhang_security_2010} developed a framework that deduces new user passwords from old user passwords through a series of successive transformations. The framework was able to infer 41\% of new passwords in an offline attack and 17\% in an online attack.  In 2014, Choong et al.~\cite{choong_united_2014}  conducted a NIST study of US government employees' password habits and found that when asked to create a new password, respondents tended to use less secure strategies, such as recycling old passwords or only making a minor change to an existing password.  

In 2015, Farcasin et al.~\cite{farcasin_why_2015} surveyed university affiliates regarding pre-generated and expiring passwords. Respondents reported that a 120-day expiration time was too short, and the authors concluded that rapid expiration is untenable for most users, leading to password reuse and the creation of less secure new passwords. In 2018, Habib et al.~\cite{habib_user_2018} also surveyed users and found that regular password replacement usually led to similarly secure new passwords. They question the security gains of an expiration policy and recommend investing into alternative security measures. 

In 2023, Gerlitz et al.~\cite{gerlitz_evolution_2023} conducted a longitudinal study of employees of three German companies about the German Federal Office for Information Security's removal of the password expiration requirement from their policy guidelines. While they reported a downward trend in requests to renew passwords, the investigation also revealed several factors that led to continued reliance of organizations on privacy expiration despite the federal recommendation against it.  Several organizations still, mistakenly, viewed the practice of periodically renewing passwords as beneficial to overall IT security. 
Several organizations also kept password expiration while transitioning to MFA or were in industries with contradictory requirements, such as finance, which continue to mandate password expiration.

\paragraph{Password Composition Rules.}
Strict password composition rules are similarly problematic. While the idea of increasing password entropy through composition rules is good in theory, real users often make predictable choices that satisfy the composition rules but leave passwords vulnerable. In 2011, Komanduri et al.~\cite{komanduri2011passwords} found that composition rules mandating a mixture of cases, numbers, and symbols resulted in lower-entropy (worse) passwords than simply mandating longer passwords with no specific composition rules. In 2015, Ur et al.~\cite{ur2015added} found that many ``weak passwords resulted from misconceptions, such as the belief that adding `!' to the end of a password instantly makes it secure or that words that are difficult to spell are more secure than easy-to-spell words.'' This misunderstanding is understandable given the widespread composition rule that passwords contain at least one symbol. Extensive composition rules also place a burden on users, resulting in less memorable passwords that are more likely  saved in an insecure location or re-used for multiple accounts.

Authentication experts and NIST SP 800-63 now support the use of password strength meters and ``increased password length\dots especially through encouraging the use of passphrases''~\cite{800-63-FAQ} instead of forced composition rules. Meters and other forms of visual feedback are less likely to lead to formulaic passwords while still giving users an indication of password strength. However, the design of the feedback matters. 
In 2008, Forget et al.~\cite{forget2008improving} proposed a system that added random characters inside user-chosen passwords to increase their entropy.
In 2017, Segreti et al.~\cite{segreti2017diversify} tested adaptive password composition policies that change over time as users create new passwords. 
In 2023, Behfar et al.~\cite{behfar2023can} explored the effectiveness of various metaphor-based password strength indicators on strong user password selection. 
In 2023, Amador et al.~\cite{amador2023prospects} found that interventions guided by prospect theory can cause users to improve password strength. 
In 2024, Paudel et al.~\cite{paudel2024priming} designed and evaluated a ``priming-through-persuasion'' approach to inform users about weak password selections.

\paragraph{Knowledge-based Authentication.}
Problems with knowledge-based authentication have been raised by the academic community for more than a decade. In 2009, Just et al.~\cite{just2009personal} found that answers to security questions were typically low entropy (easy to crack) and users had trouble remembering their answers. In 2015, Bonneau et al.~\cite{bonneau2015secrets} examined personal knowledge questions at Google and found ``a security level that is far lower than user-chosen passwords'' with many users providing difficult-to-remember fake answers, concluding that ``best practice should favor more reliable alternatives.''

\paragraph{Authentication Security in Higher Education.}
A few studies have examined other aspects of authentication security in higher education. In 2022, Mayer et al.~\cite{mayer2022users} studied why faculty, staff, and students at large educational institutions chose to use password managers or not. They found that perceived ease of use was the most important factor and recommended advocacy focusing on usability benefits. 
In 2023, Nisenoff et al.~\cite{nisenoff2023two} found that many university accounts were vulnerable to credential-guessing attacks performed using cracked passwords from a data breach matched with email addresses. 

\paragraph{Compliance with NIST \textit{Digital Identity Guidelines}.}
Relatively few studies have examined compliance with the 2017 NIST SP 800-63 \textit{Digital Identity Guidelines}. 
In 2022, Lee et al.~\cite{lee2022password} ``examined the [password] policies of 120 of the most popular websites'' and ``found that only 13\% of websites followed all relevant best practices\ldots75\% of websites do not stop users from choosing the most common passwords\ldots45\% burden users by requiring specific character classes in their passwords for minimal security benefit.''
In 2023, Hall et al.~\cite{hall2023empirical} analyzed over 100 websites across industries that ``report  the  most  breaches  in  the  Verizon  Data  Breach  Investigation  Report.'' They found a mixture of compliance and noncompliance, including ``nearly all websites\ldots avoiding the use of security questions and SMS-based 2FA'' (a  higher compliance rate than the institutions in our study), but that ``many websites (greater than 80 percent) still deem `P@ssw0rd' an acceptable password.''
Our work is the first to examine the compliance of higher education institutions with NIST SP 800-63 \textit{Digital Identity Guidelines}  specifically and at scale. 

%% file: sections/methods.tex
\section{Methods}
\label{sec:methods}

This section presents our methods, including which higher education institutions we chose to investigate and why, how we reviewed publicly available online documentation from these institutions for details about the authentication policies of interest (Section~\ref{sec:background}), and some limitations that may arise from the nuances of our methods.

\subsection{Institution Selection}

There are over 5500 higher education institutions in the United States~\cite{nces_school_enrollment} and over 400 in Canada~\cite{cmec_education_in_canada}. Due to resource constraints that made investigation of the entire higher education sector infeasible, we selected a limited set of institutions as the focus of this study. It has been well observed that the actions of a few well-regarded institutions often have an outsize influence on behavior across the higher education space. We therefore adopted a selection process that prioritized well-regarded institutions from several broad categories.

Specifically, we selected institutions from the U.S. News and World Report lists of ``top’’ institutions from several categories as of September 2023. We chose the top 20 Canadian global universities \cite{usnews_global_universities_canada_2023}, top 20 US national universities \cite{usnews_best_colleges_2023}, top 20 US public colleges and universities \cite{usnews_public_2023}, top 20 US liberal arts colleges \cite{usnews_best_liberal_arts_colleges_2023}, top 10 US historically Black colleges and universities (HBCUs) \cite{usnews_hbcu_2023}, and a selection of US regional colleges consisting of the top 10 in each of the North, South, Midwest, and West regions\footnote{Less one institution  that closed during the study (University of Antelope Valley).} \cite{usnews_regional_colleges_2023}. Since multiple institutions may tie for a ranking on these lists, this selection process resulted in \num total institutions (Table~\ref{tab:institutions}).
We recognize that this particular list of institutions is not without limitations, which we discuss in greater detail in Section~\ref{sec:limitations}.

\input{tables/universities-table-results}

\subsection{Codebook Preparation}
\label{sec:codebook}

Since each of the NIST standards we investigated has a ``should'' (``shall'') or ``should not'' format, we used a deductive coding approach with a fixed set of codes determined prior to data collection.
We chose a codebook applicable to all of the standards that would be easily understood by trained undergraduate researchers. 
Each of the six codes were potential answers to the following question: ``Based on publicly available online documentation, what is the institution's policy regarding the practice associated with this standard?''

\begin{enumerate}
    \item ``Required'': The practice is required for all affiliates of the institution.
    \item ``Required for Specific Affiliates'': At least some affiliates of the institution are required to follow the practice. In this case, we also recorded which specific affiliates are subject to this requirement. 
    \item ``Recommended'': Affiliates of the institution are recommended to follow the practice, but adherence is not enforced.
    \item ``Discouraged'': Affiliates of the institution are discouraged from following the practice, but prevention is not enforced. 
    \item ``Disallowed'': Affiliates of the institution are not permitted to follow the practice.
    \item ``No Information Found'': The institution's publicly available online documentation does not provide any or enough information to determine the institution's policy regarding the practice.
    \end{enumerate}

We confirmed the effectiveness of this codebook through trial data collection with three institutions. The codes covered all cases appearing in the trial. 

\subsection{Data Collection}

This section describes the data collection process followed for all pairs of investigated institutions and standards.

\subsubsection{Policy Identification}

Institutions are inconsistent as to how they post information about authentication policies online.
We therefore used several methods to locate publicly available policies on a wide variety of official institutional websites, including (but not limited to): 

\begin{enumerate}
\item Search engines (e.g., Google) with site-specific search queries (``site:[institution].edu'')
\item Search bars on institutional websites
\item Manual navigation of institutional websites (e.g., looking for links to policies or documents about digital authentication)
\end{enumerate}

This process was intentionally open-ended, reflecting the variety of institutional website structures and information posting practices. 
Relevant policies most frequently appeared on 
IT support pages (also referred to as IT ``guides,'' ``help desks,'' or ``knowledge bases''), information security policies, cybersecurity policies, and IT policies.

If we found multiple policies for a particular institution, we selected institution-wide policies over policies for individual academic departments or administrative units. 
If we found multiple institution-wide policies from different years, we selected the newest policy at the time of data collection.
We excluded policies from websites clearly indicated as over 10 years old.

\subsubsection{Policy Coding}
We read the identified policies, selected the corresponding code from the codebook for each investigated standard,
and recorded a link to the policy website to facilitate reproducibility.
For the password expiration standard, we also recorded required or recommended password expiration frequencies. For the multi-factor authentication standard, we also recorded what third-party MFA provider was supported or recommended.

\subsubsection{Researcher Roles}

The policy identification and coding processes described above were performed independently by two non-overlapping sets of researchers:

\begin{enumerate}
\item The second author initially performed policy identification and coding for all institutions and standards in the study. Their work was then reviewed by a law student research assistant, who visited all recorded links and highlighted any codes they believed were incorrect. The third and fourth authors resolved all disagreements and selected final codes.
\item A group of 25 undergraduate computer science students performed policy identification and coding for all institutions and standards as a part of a class on cybersecurity. These students were trained by the first author, and each student collected data and coded policies for 1--6 institutions. 
\end{enumerate}
The first author then reviewed and merged all codes produced by (1) and (2) above, resolving disagreements, updating codes for websites that had changed between the initial coding and May 2025, and recording screenshots or PDFs of the institutional policies for reproducibility. 

\subsection{Reproducibility}
We have posted a reproducibility package in a GitHub repository: \url{https://github.com/NoahApthorpe/apthorpe-soups-2025-data}. The repository contains a CSV file listing the institutions, institution categories, links to policy websites, all codes, and all additional information collected for the four authentication standards analyzed in this paper.
The repository also contains folders with screenshots or PDF documents from the policy websites that were taken in May 2025. Note that the results in this paper reflect these screenshots and PDF documents, although the institutions' live websites may have changed since these data were collected.

\subsection{Research Ethics}
The Colgate University Institutional Review Board determined that this study was exempt. We only report data from publicly available sources posted online by the institutions we study. 
We do not collect or report any information about individual affiliates at any institution. 

All contributing undergraduate computer science students were informed ahead of time that their work may be included in a publication. Their involvement occurred during a normally scheduled lab period and was incorporated into course learning goals. The students received course participation credit for their involvement and are thanked in the acknowledgments.

\subsection{Limitations}
\label{sec:limitations}

The methods used in this paper have some limitations that should be acknowledged to understand the scope of the results.

First, it is essential to remember that this study examines institutional \textit{policies}, and that these policies may or may not represent actual, technologically-enforced institutional practices. For example, a policy might say that multi-factor authentication is required for all institution affiliates, but some affiliates might, in practice, be able to disable MFA via their internal profile settings. Similarly, a policy might state that yearly password expiration is required for all affiliates, but the current IT leadership might not actually enforce regular password cycling. Nevertheless, we expect posted policies to generally align with actual practices, especially since posted policies are often primarily intended as guides for affiliates (e.g., on IT ``help'' websites). The policies also signal the intentions and values of the institution and may influence the actions of affiliates and peer institutions.  

On a practical level, determining actual technically-enforced practices would require institutional affiliation or direct input from institutional representatives, which would be challenging.
We attempted to address this limitation by emailing the IT departments at 100 of the institutions included in the study with requests to confirm or correct the information we had collected about their policies. However, we only received four meaningful responses, and those effectively corroborated our existing data. We received nine additional responses claiming that no additional data could be provided. 

Second, the \num institutions we studied represent only a fraction of the higher education sector. We encourage follow-up studies examining authentication practices and other forms of cybersecurity compliance across other countries\cite{mayer2017second} and institutional categories. The institutions in this paper are at the top end of prestige in their respective categories, which must be remembered when interpreting the results. These institutions, in contrast to less prestigious peers, are more likely to be well-resourced, making them more likely to have sufficient staff with expertise and availability to ensure the institution’s authentication (and other cybersecurity) practices follow accepted guidelines. The fact that we find many cases of noncompliance with NIST standards among these prestigious institutions’ online policies suggests that the overall rate of noncompliance is substantially higher across the higher education space. 

Third, this study reflects a snapshot in time and only a subset of the authentication standards in NIST SP 800-63. Follow-up longitudinal studies are necessary to see whether the trajectory of institutional policies and practices is trending toward compliance with these and other NIST standards. 

%% file: tables/universities-table-results.tex
\begin{table*}[ht]
\tiny
\centering
\begin{tabular}{@{\,\,\,}l@{\,\,\,}c@{\,\,\,}c@{\,\,\,}c@{\,\,\,}c@{\qquad}l@{\,\,\,}c@{\,\,\,}c@{\,\,\,}c@{\,\,\,}c@{\qquad}l@{\,\,\,}c@{\,\,\,}c@{\,\,\,}c@{\,\,\,}c@{\,\,\,}}
\toprule
US National Universities & MFA & PE & CR & KBA & US Public Colleges and Universities & MFA & PE & CR & KBA & US Liberal Arts Colleges & MFA & PE & CR & KBA \\
\midrule
Princeton University & A & N & X & N & University of Michigan--Ann Arbor & A & N & N & X & Williams College & A & S & A & N \\
Massachusetts Institute of Technology & A & A & A & N & University of North Carolina at Chapel Hill & A & A & S & A & Amherst College & A & S & A & N \\
Harvard University & A & N & R & N & University of Virginia & A & A & A & R & United States Naval Academy & N & A & A & A \\
Stanford University & A & N & S & A & University of California, Davis & A & N & N & A & Pomona College & A & X & A & N \\
Yale University & A & A & S & X & University of California, San Diego & A & N & S & A & Swarthmore College & A & A & R & N \\
University of Pennsylvania & A & X & S & N & University of Florida & A & A & A & N & Wellesley College  & A & R & A & N \\
California Institute of Technology & A & N & R & R & University of Texas at Austin & A & N & A & N & United States Air Force Academy & A & N & A & N \\
Duke University & A & X & X & N & Georgia Institute of Technology & A & A & A & A & United States Military Academy at West Point & A & N & S & N \\
Brown University & A & R & A & R & University of California, Irvine & A & R & A & A & Bowdoin College & A & N & A & A \\
Johns Hopkins University & A & A & A & A & University of California, Santa Barbara & A & R & R & A & Carleton College & A & S & S & N \\
Northwestern University & A & A & A & A & University of Illinois Urbana-Champaign & A & A & A & N & Barnard College & A & N & N & N \\
Columbia University & A & D & A & A & University of Wisconsin--Madison & A & R & R & A & Claremont McKenna College & A & R & R & S \\
Cornell University & A & N & A & D & Rutgers University--New Brunswick & A & N & N & A & Grinnell College & S & N & S & N \\
University of Chicago & A & S & S & A & University of Washington & A & N & A & D & Middlebury College & A & N & A & N \\
University of California, Berkeley & A & N & A & S & The Ohio State University & A & A & A & A & Wesleyan University & A & S & N & N \\
University of California, Los Angeles & A & N & R & N & Purdue University--Main Campus & A & A & A & A & Davidson College & R & N & X & N \\
Rice University & A & R & A & N & University of Maryland, College Park & A & A & A & X & Hamilton College & A & X & X & A \\
Dartmouth College & A & D & X & X & Texas A\&M University & S & S & A & R & Harvey Mudd College & A & N & A & R \\
Vanderbilt University & A & A & N & R & University of Georgia & A & S & A & A & Smith College & A & R & N & S \\
University of Notre Dame & A & X & N & A & Virginia Tech & A & A & S & N & Vassar College & A & N & N & N \\
 &  &  &  &  &  &  &  &  &  & Colgate University & A & N & R & N \\
 &  &  &  &  &  &  &  &  &  & Haverford College & A & N & N & N \\
 &  &  &  &  &  &  &  &  &  & Washington and Lee University & A & X & A & N \\
 
 \midrule
Canadian Global Universities & MFA & PE & CR & KBA & US Historically Black Colleges and Universities & MFA & PE & CR & KBA & US Regional Colleges -- North & MFA & PE & CR & KBA \\
\midrule

University of Toronto & A & N & R & R & Spelman College & R & N & A & N & United States Coast Guard Academy & N & N & S & N \\
University of British Columbia & A & A & R & A & Howard University & A & N & N & A & Cooper Union for the Advancement of Science and Art & A & R & R & N \\
McGill University & A & N & N & R & Florida A\&M University & A & A & A & A & United States Merchant Marine Academy & N & A & N & N \\
University of Alberta & S & N & A & N & Tuskegee University & N & R & A & N & Grove City College & N & N & N & N \\
McMaster University & A & R & A & N & Morehouse College & A & S & N & N & Maine Maritime Academy & S & A & A & N \\
Universite de Montreal & N & N & A & N & Xavier University of Louisiana & A & N & A & N & Pennsylvania College of Technology & S & N & N & N \\
University of Calgary & A & N & R & N & Hampton University & S & S & N & S & Elmira College  & R & N & N & N \\
University of Waterloo & A & R & X & S & North Carolina A\&T State University & S & R & R & N & Alfred State College--SUNY & S & A & A & R \\
University of Ottawa & A & A & A & N & Delaware State University & N & A & A & A & College of Mount St. Vincent & A & N & A & N \\
Western University  & A & R & A & S & Morgan State University & A & A & N & A & SUNY College of Technology at Canton & S & R & R & N \\
Dalhousie University & A & N & X & N &  &  &  &  &  & University of Maine at Farmington & R & A & A & A \\
Simon Fraser University & A & A & S & N &  &  &  &  &  &  &  &  &  &  \\
University of Victoria & A & N & N & N &  &  &  &  &  &  &  &  &  &  \\
University of Manitoba & A & A & A & R &  &  &  &  &  &  &  &  &  &  \\
Laval University & A & N & A & N &  &  &  &  &  &  &  &  &  &  \\
York University - Canada & A & A & R & R &  &  &  &  &  &  &  &  &  &  \\
Queens University - Canada & A & A & R & A &  &  &  &  &  &  &  &  &  &  \\
University of Saskatchewan & A & N & A & R &  &  &  &  &  &  &  &  &  &  \\
University of Guelph & A & R & A & R &  &  &  &  &  &  &  &  &  &  \\
Carleton University & A & N & A & R &  &  &  &  &  &  &  &  &  &  \\

\midrule
US Regional Colleges -- South & MFA & PE & CR & KBA & US Regional Colleges -- Midwest & MFA & PE & CR & KBA & US Regional Colleges -- West & MFA & PE & CR & KBA \\
\midrule

High Point University & A & N & N & N & Illinois Wesleyan University & A & N & A & N & Embry-Riddle Aeronautical University--Prescott & S & N & R & A \\
Florida Polytechnic University & A & A & N & N & Ohio Northern University & N & N & N & N & California State University--Maritime Academy & A & A & A & N \\
Beacon College & S & S & S & N & College of the Ozarks & N & N & N & N & Carroll College & N & N & N & N \\
Flagler College & A & S & A & A & Taylor University & S & N & A & N & Criswell College & S & N & N & N \\
Wesleyan College & S & N & A & A & Simpson College & N & N & A & N & Oregon Institute of Technology & N & A & A & A \\
Catawba College & A & N & A & A & Cottey College & N & N & N & N & Fashion Institute of Design \& Merchandising & N & N & N & N \\
University of the Ozarks & N & N & N & N & Alma College & S & A & S & N & College of Idaho & A & A & A & N \\
Spring Hill College & N & A & N & A & Benedictine College   & A & A & A & N & Brigham Young University--Hawaii & A & N & R & A \\
Huntingdon College & N & N & N & N & William Jewell College & R & S & A & N & Brigham Young University--Idaho & A & R & N & R \\
Barton College & N & R & N & N & Hiram College & A & S & A & A &  &  &  &  &  \\
Newberry College & N & S & A & N & Lake Superior State University & N & N & N & N &  &  &  &  &  \\
\bottomrule
\end{tabular}

\captionsetup{font=small}
\caption{
Publicly available policies regarding the four authentication standards from NIST SP 800-63 investigated in this study: MFA = multifactor authentication, PE = password expiration, CR = password composition rules, KBA = knowledge-based authentication.
Code abbreviations refer to the codebook in Section~\ref{sec:codebook}: A = required for all affiliates, S = required for specific affiliates, R = recommended, D = discouraged, X = disallowed, N = no information found.
These institutions were selected from the 2023 U.S. News and World Report lists of top institutions in their respective categories \cite{usnews_best_colleges_2023, usnews_best_liberal_arts_colleges_2023, usnews_regional_colleges_2023, usnews_global_universities_canada_2023, usnews_public_2023, usnews_hbcu_2023}. 
Additional information, including specific affiliates, password expiration frequencies, policy links, and policy screenshots or PDF documents are available at \url{https://github.com/NoahApthorpe/apthorpe-soups-2025-data}.}
\label{tab:institutions}
\end{table*}

%% file: sections/results.tex
\section{Results}
\label{sec:results}

Our results for the four standards and \num institutions we investigated are shown in Table~\ref{tab:institutions}. Additional information, including specific affiliates, password expiration frequencies, policy links, and policy screenshots or PDF documents are available in the following GitHub repository: \url{https://github.com/NoahApthorpe/apthorpe-soups-2025-data}.
This section summarizes these results.

\begin{figure*}[tp]
    \centering
    \includegraphics[width=0.45\linewidth]{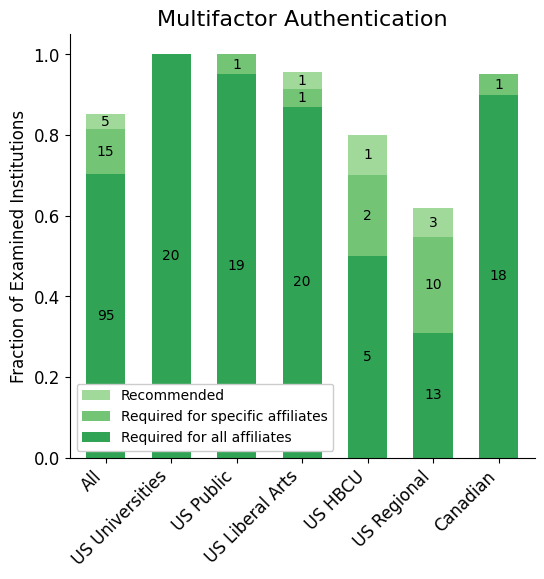} \quad
    \includegraphics[width=0.45\linewidth]{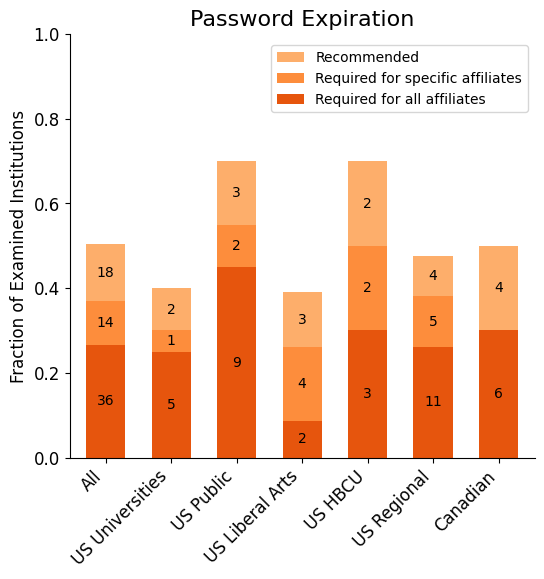}
    \includegraphics[width=0.45\linewidth]{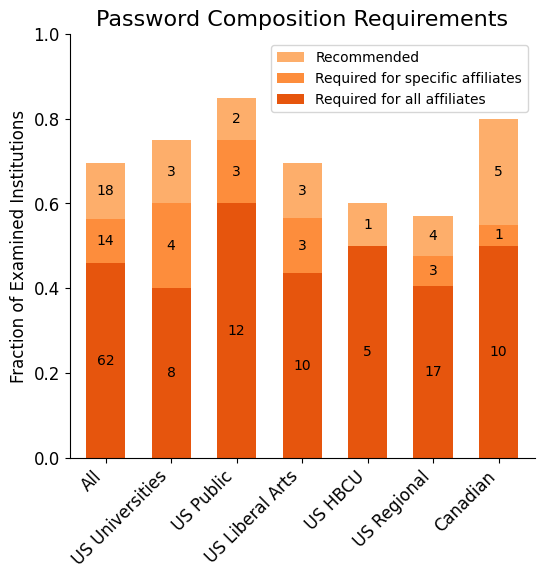} \quad
    \includegraphics[width=0.45\linewidth]{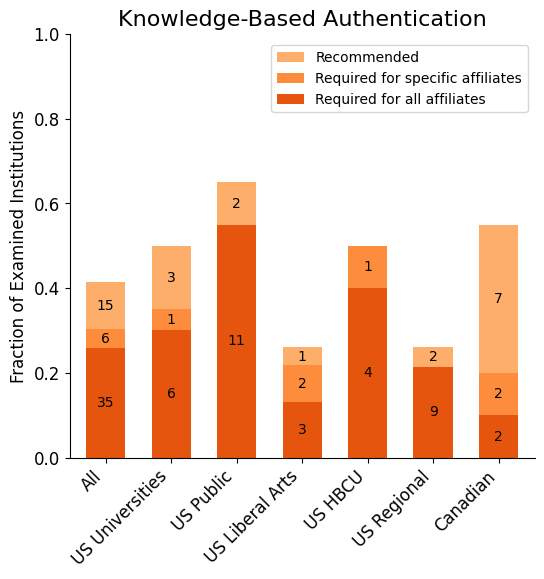}

    \caption{Online publicly available policies of \num higher education institutions regarding standards from NIST SP 800-63. Policies about multi-factor authentication (in green) align with the standards. Policies about password expiration, password composition requirements, and knowledge-based authentication (in red) are noncompliant.}
    \label{fig:results}
\end{figure*}

\subsection{Multi-factor Authentication}

The use of MFA is widespread in higher education (Figure~\ref{fig:results}). 70\% (95) of all investigated institutions require MFA for all affiliates according to their online policies. 11\% (15) require MFA for some affiliates, and 4\% (5) at least recommend MFA. 
The institutions that require MFA  at least for some affiliates do so for students, employees, remote users, privileged accounts, or accounts with access to ``sensitive'' data. Most of the institutions requiring or recommending MFA use a third-party service, such as Duo or Microsoft Authenticator, to provide MFA functionality.

At least 90\% of investigated US national universities, US public colleges and universities, Canadian global universities, and US liberal arts colleges require MFA for some or all affiliates. 
A considerably lower percentage of US HBCUs and US regional colleges require MFA than institutions in other categories (70\% and 55\% versus $\ge$90\%, respectively). 

\subsection{Password Expiration}

Requiring or recommending password expiration/cycling is still common across higher education (Figure~\ref{fig:results}). 27\% (36) of all investigated institutions require regular password expiration for all affiliates,  10\% (14) require regular password expiration for specific affiliates (e.g., faculty and staff, students, affiliates with passwords shorter than a threshold length, or affiliates without MFA enabled), and 13\% (18) recommend (but do not require) regular password expiration. A larger percentage of US public colleges and universities (55\%) and US HBCUs (50\%) require password expiration than institutions in other categories ($\le$40\%). 

One year was the most common password expiration frequency we observed, but specific frequencies ranged from expiration every month to expiration frequencies that vary depending on whether an affiliate has enabled MFA or whether the affiliate is an administrator or a regular user (Table~\ref{fig:expiration-times}). Some institutions did not list an exact expiration frequency, rather recommending password expiration ``periodically,'' ``often,'' or ``frequently.''
These policies do not comply with the NIST 800-63 standard that password expiration should only occur in response to a known breach.

\begin{table}[t]
\small
\centering
\begin{tabular}{lr}
\toprule
Password Expiration Frequency & Count \\
\midrule
1 year (annually, 365 days, etc.) & 32 \\
180 days & 6 \\
90 days & 6 \\
6 months & 5 \\
Periodically & 5 \\
60 days & 2 \\
Often & 2 \\
1 year (user), 180 days (admin) & 1 \\
1 year (with MFA), 180 days (without MFA) & 1 \\
1 year (with MFA), 120 days (without MFA) & 1 \\
180 days (user), 90 days (admin) & 1 \\
126 days & 1 \\
120 days & 1 \\
30 days & 1 \\
30 days (sensitive data), 90 days (other data) & 1 \\
Once or twice a year & 1 \\
Frequently & 1 \\
\bottomrule
\end{tabular}
\caption{Required or recommended password expiration frequencies in the online policies of the investigated institutions.}
\label{fig:expiration-times}
\end{table}

Some of these institutions purposefully disregard the NIST standard. For example, one institution has a Q\&A on ``Why Are We Implementing A Password Change Initiative When It Is No Longer Considered Best-Practice?'' that states
\begin{quote}
Changing passwords regularly and implementing a password expiry date helps to limit the use of compromised accounts by attackers for malicious activities. In an effort to provide better account management while adhering to the spirit of best practice guidance, a reset interval that is longer than “90 days” but shorter then “never” is being put into practice. 
\end{quote}

\subsection{Password Composition Rules}

\begin{figure}[t]
\centering
\includegraphics[width=0.75\linewidth]{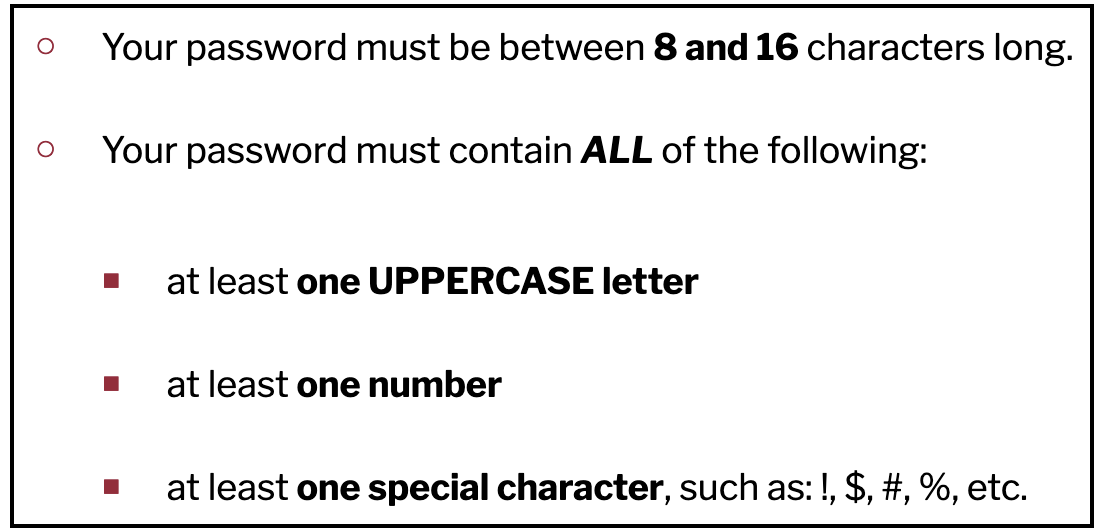}
\caption{Example password composition rules from a US regional college.} 
\label{fig:comp-rule}
\end{figure}

Despite the NIST SP 600-83 standard that institutions should not impose  password composition rules because they ``do not significantly improve the security of selected passwords''~\cite{800-63-FAQ}, 
the presentation and enforcement of password composition rules remains widespread across higher education (Figure~\ref{fig:results}). 46\% (62) of all investigated institutions require that all affiliates meet minimum password composition rules during the password creation process. 10\% (14) require that at least some affiliates meet password composition rules, including administrative accounts, accounts with MFA not enabled, or accounts with passwords shorter than a threshold length. 
Another 13\% (18) recommend (but do not require) that affiliates meet password composition rules. 
Figure~\ref{fig:comp-rule} shows a representative example of password composition rules from a US regional college.

Rates of password composition requirements are relatively consistent across institution categories. A somewhat higher percentage of US public colleges and universities (75\%) have password composition requirements than institutions in other categories ($\le 60\%$). 

\subsection{Knowledge-based Authentication}

The use of security questions (e.g., ``What is the name of your first pet?'') during the authentication process remains common across higher education (Figure~\ref{fig:results}). 26\% (35) of all investigated institutions require security questions for all affiliates, 4\% (6) require security questions for some affiliates, and an additional 11\% (15) recommend (but do not require) security questions as an authentication option. 
All such policies are out of compliance with the NIST SP 800-63 standard that security questions are no longer recommended for any authentication process due to their inherent insecurity.
Overall, a larger percentage of US public colleges and universities (55\%) and US HBCUs (40\%) require security questions than institutions in other categories ($\le$30\%).
 
The institutions that require security questions typically do not provide any guidance about which questions or what types of answers are likely to be more secure. One institution described transitioning from requiring one security question to requiring three security questions. While this approach may be slightly more secure, it is still out of compliance with the NIST standard.

%% file: sections/discussion.tex
\section{Discussion}
\label{sec:discussion}

This section discusses the implications of our results, emphasizing insights for the usable security community.

\paragraph{Cyberattack Vulnerability.}

The increased use of remote learning and e-learning platforms has increased the threat surface of higher education~\cite{bandara2014cyber}. According to Alexei et al.~\cite{alexei2021cyber}, the threat of ``DoS / DDoS attacks, cross-site scripting, spoofing, unauthorized data access and infection with malicious programs, [and] also the theft of personal data has increased dramatically.'' Ransomware~\cite{serna2022increase, suarez2024unsafe, acosta2023empirical} and phishing~\cite{broadhurst2019phishing, diaz2020phishing, alkhalil2021phishing} have also become more common, with ransomware attack prevalence increasing over 70\% in 2023~\cite{cozens2024ransomware} and recovery costs averaging around \$1 million~\cite{schwartz2023ransomware}.

Given these conditions, the widespread deployment of MFA is encouraging, as the requirement of a second factor substantially decreases the risk of credential-based attacks. However, the persistence of institutions not requiring MFA (especially regional colleges) and  the continued use of noncompliant password policies and knowledge-based authentication leaves the sector vulnerable. The NIST SP 800-63 compliance rates we observe for higher education institutions are lower than those observed by Hall et al.~\cite{hall2023empirical} across a swath of industries identified as prone to data breaches. Our study should remind higher education IT departments and administrations to require MFA and update outdated password and KBA policies. 

\paragraph{Limited Policy Disclosure.} The availability of online documentation about authentication practices varied widely across the institutions we examined. This means that the rates of noncompliance we observe may be a lower bound, with other noncompliant institutions potentially choosing not to post information about their authentication practices online.

Digital authentication policies should be readily available, periodically updated to reflect current practices, and standardized to clearly communicate relevant information. Disclosure of such policies could motivate other organizations to follow similar practices and serve an educational purpose, teaching affiliates about the authentication practices they should understand and expect in other aspects of their online lives. We hope that our research incentivizes more institutions to bring their digital authentication practices into compliance with known best practices and provide public disclosure that they have done so. 

To simplify this process, we advocate for the standardization of public-facing authentication policy notices, e.g., via a free template. This would make it simple for institutions to see what changes are necessary to bring their practices into compliance and to quickly create a user-friendly document describing their practices with a widely-recognized and ideally machine readable format. Such a template should allow easy customization for details unique to an institution within constraints ensuring that the resulting notice still describes compliant practices. 
The specifics of this standardization would need to be developed in conversation with higher education institutions, but preliminary versions could be proposed and tested in future usable security research. 
This would be similar to prior usable security work on the standardization of privacy policies~\cite{kelley2010standardizing, cranor2012necessary}.

\paragraph{Data Confirmation Difficulty.}

The difficulty we faced determining whether actual authentication practices match online policies (Section~\ref{sec:limitations}) shows that  institutions are often unwilling or unable to share details of cybersecurity practices with outside researchers, a phenomenon echoed in the literature~\cite{botta2007studying, kotulic2004there}. One institution's IT security director even contacted us to apologize that while they would have liked to share more details about their institution's practices, they had been forbidden to do so by those higher in administration.  

However, the secrecy of details at the level of the NIST standards are irrelevant to the security of an authentication system  (see Kerckhoffs's principle~\cite{petitcolas2023kerckhoffs} and the ineffectiveness of ``security through obscurity'').
The very same institutions that balk at contributing to academic research vetted by institutional review boards readily contract with private education technology platforms having known privacy and security vulnerabilities~\cite{sanfilippo2023privacy}. 
We hope that the resistance towards sharing cybersecurity data with academic researchers abates with continued advocacy. 

\paragraph{Explaining Noncompliance: Incentives.}

The noncompliance we observe could be understood strictly in terms of incentives and resources. From this perspective, the NIST standards constitute non-binding, technical guidance about authentication and digital identity that informs cost-benefit calculations, risk assessment/management, procurement, and other organizational decision-making processes. Higher education organizations are complex, and various departments and personnel may be responsible for taking NIST standards into account. Resource constraints play a substantial role in cybersecurity~\cite{moodys2023cybersurvey}, and variations in resources across institutions likely impact (non)compliance with NIST standards.
Entrenched legacy systems~\cite{althani4783029technical}, difficult to configure third-party software (e.g., single sign-on platforms~\cite{grinich2022sso}), staff shortages~\cite{crumpler2022cybersecurity}, and institutional prioritization of features other than cybersecurity could all contribute to resource shortages and incentives leading to noncompliance.

To increase the rate of compliance and consequently improve cybersecurity in higher education, society might need to re-engineer the availability of relevant resources {and thereby change the corresponding incentives. This might entail increasing public funding directed towards better cybersecurity practices (e.g., to support MFA), improving disclosure of policies and practices (which might impact accountability to different stakeholders), or adjusting incentives and priorities through external pressure (e.g., insurance, markets, even legal reform).

However, this perspective does not provide a fully satisfactory explanation for persistent noncompliance with NIST standards that would otherwise reduce burden on users, decrease regular IT support costs, and improve authentication security. Password expiration is the most obvious example. Refraining from  regular password replacement, except when a breach has occurred, reduces users' time spent changing passwords and users' mental effort remembering (or saving) new passwords. After initial system configuration updates, IT staff are relieved from recurring technical password expiration processes and from user support for lost or incorrect new passwords. Based on these aligned incentives, compliance should be widespread, yet our findings suggest it is not. 
This suggests that the noncompliance we observe is at least partially due to other factors, such as 
knowledge commons problems.

\paragraph{Explaining Noncompliance: Knowledge Diffusion.}
Knowledge commons problems~\cite{frischmann2014governing} concern the \textit{diffusion} of evolving expert knowledge, i.e., how such evolving knowledge translates (or not) into professional practice and system design.
Based primarily on interviews of cybersecurity experts, Frischmann and Johnson \cite{frischmann2023common} have identified concerns that might impede expert knowledge diffusion, including that ``professionals may be overconfident and inaccurately see themselves as security experts or as being sufficiently up-to-date on security.''

If the noncompliance we observe is indeed a consequence of a delayed diffusion of expert knowledge to many higher education institutions, it may work itself out over time as more practitioners gradually become aware of the updated standards. 
NIST itself has number of practices to facilitate the dissemination of its guidance, including public websites~\cite{nist_csrc}, direct engagement with universities and colleges~\cite{nist_educational_outreach}, cybersecurity clinics~\cite{nist_cybersecurity_clinics_2025}, and mailing lists~\cite{nist_fissea_mailing_list}.

However, given the length of time since the NIST SP 800-63 \textit{Digital Identity Guidelines} were published, our results suggest that new forms of engagement may be necessary to facilitate the timely diffusion of expert knowledge to those actually responsible for setting and implementing authentication policies and practices. 
There is an extensive usable security literature about the diffusion of software patches and other updates to users, system administrators, and network administrators \cite{jenkins2024not, mathur2016they, dissanayake2022and, li2019keepers, tiefenau2020security} that could inform new methods of overcoming knowledge diffusion roadblocks in the cybersecurity standards space. 
Qualitative interviews of IT professionals at higher education institutions could also provide insight into overlooked or misunderstood reasons for persistent noncompliance.  

We hope that the usable security community takes up this challenge, as a combination of human-computer interaction and sociotechnical systems research will be necessary to monitor and improve the rapid adoption of new best practices espoused in future standards updates.

%% file: sections/conclusion.tex
\section{Conclusion}
\label{sec:conclusion} 

We examined online policies of \num higher education institutions in the US and Canada to measure compliance with NIST SP 800-63 \textit{Digital Identity Guidelines}. We focused on four standards that reflect changes from prior versions and impact all institutional account holders. We found widespread, but not universal, deployment of multi-factor authentication. We also found widespread noncompliance with standards for password expiration, password composition rules, and knowledge-based authentication.  
These results indicate that while some expert cybersecurity recommendations have substantively influenced the policies of higher education institutions, more investment and outreach in this area is needed, as well as more research into incentive structures and the diffusion of expert guidance to practitioners.